\documentclass[aps,pre,twocolumn,groupedaddress]{revtex4}
\usepackage{graphicx}
\usepackage{dcolumn}
\usepackage{bm}
\usepackage{amssymb}
\usepackage[colorlinks,
                   linkcolor=blue,
                  anchorcolor=blue,
                  citecolor=blue,
                  urlcolor=blue,
                  CJKbookmarks=true
                  ]{hyperref}

\begin{document}


\title{Theoretical modelling discriminates the stochastic and deterministic hypothesis of cell reprogramming}


\author{Jiawei Yan}
\email[]{yann.jiawei@gmail.com}
\altaffiliation{School of Life Sciences, Peking University, Beijing 100871, China}
\author{Pu Zheng}
\altaffiliation{School of Life Sciences, Peking University, Beijing 100871, China}
\author{Xingjie Pan}
\altaffiliation{School of Physics, Peking University, Beijing 100871, China}

\begin{abstract}
How to induce differentiated cells into pluripotent cells has elicited researchers' interests for a long time since pluripotent stem cells are able to offer remarkable potential in numerous subfields of biological research. However, the nature of cell reprogramming, especially the mechanisms still remain elusive for the sake of most protocols of inducing pluripotent stem cells were discovered by screening but not from the knowledge of gene regulation networks. Generally there are two hypotheses to elucidate the mechanism termed as elite model and stochastic model which regard reprogramming process a deterministic process or a stochastic process, respectively. However, the difference between these two models cannot yet be discriminated experimentally. Here we use a general mathematical model to elucidate the nature of cell reprogramming which can fit both hypotheses. We investigate this process from a novel perspective, the timing. We calculate the time of reprogramming in a general way and find that noise would play a significant role if the stochastic hypothesis holds. Thus the two hypotheses may be discriminated experimentally by counting the time of reprogramming in different magnitudes of noise. Because our approach is general, our results should facilitate broad studies of rational design of cell reprogramming protocols.
\end{abstract}


\maketitle
\section{Introduction}
\label{intro}
 Embryonic stem cell, which is able to divide indefinitely while maintaining pluripotency, is expected to be useful in numerous subfields of life sciences, including toxicology, disease mechanism research\cite{I1} and regeneration medicine\cite{I2}\cite{I3}. In 2006, Yamanaka and his colleagues achieved artificially induced pluripotent stem (iPS) cells from mouse fibroblasts\cite{I4}, and later from adult human fibroblasts\cite{I41}.

 Induction of iPS cell is in a very low efficiency ($\sim0.05\%$) at first\cite{I4}\cite{I41}, and there are several models to explain the low efficiency\cite{I42}. One of these models, the stochastic hypothesis, suggests that most or all the cells are competent for reprogramming but limited of randomness, only a part of cells are able to be reprogrammed in one experiment. On the other hand, elite model presupposes that only a few cells in a culture are competent for reprogramming and the process is deterministic. Generally the process of cell reprogramming was regarded as a stochastic process\cite{I5}\cite{I6} at first. There are numerous experiments indicated that the process of cell reprogramming consists of an early stochastic phase and a late hierarchic phase. The low efficiency also implies there is stochasticity. From the point of control theory, this model may be elucidated as a core control part, which is multistable and transits from one steady state to another one stochastically, and a downstream signal transduction cascade\cite{I6}. However, these two models are soon challenged by other research which shows that cell reprogramming is a deterministic process\cite{I7} and especially almost all the cells are able to be reprogrammed into iPS cells, which conflicts with the elite model. The debate of whether the nature of cell reprogramming process is stochastic or deterministic hasn't ended yet, since the gene regulation networks of cell differentiation are still poorly understood\cite{MacArthur_2009}, thus it's hard to predict how to induce iPS cells efficiently. Although the biological experiments have yet reveal the nature of cell reprogramming, there are other approaches to investigate this question. Because the elite model supposes only a fraction of cells can be reprogrammed, and this paper only focus on single cell dynamics, we use the term 'deterministic hypothesis' instead of 'elite model' to represent the hypothesis that cell reprogramming is a deterministic process.

 Our aim is to discriminate these two hypotheses, the stochastic hypothesis and deterministic hypothesis, especially in the way that the experiments can apply to. These two hypotheses, apparently, are different since one model suppose that not all the cells can be reprogrammed statistically while the other one does. However, this difference isn't able to be tested so far by experiments, because the efficiency of reprogramming may not only determined by the nature of reprogramming but also by the other factors including the environment and the method itself. Therefore, we turned to the dynamics of reprogramming. From a perspective of nonlinear dynamics, different cell fates are able to be regarded as a steady state of a dynamical system, which is the core gene regulation network for cell fate decision. This concept was noted by C. Waddington in the 1940s, which by now is called Waddington's epigenetic landscape \cite{Waddington}. In the Waddington's metaphor, the cell fate is like a ball and the landscape epitomizes the constraints of dynamical system, thus the cell fate can transit by stochastically 'jump' from one valley to another and the probability of transition depends on the magnitude of noise and the height of the hill between two valleys (the energy barrier)\cite{MacArthur_2009}\cite{Enver}.

\begin{figure}
 \resizebox{0.5\textwidth}{!}{%
\includegraphics{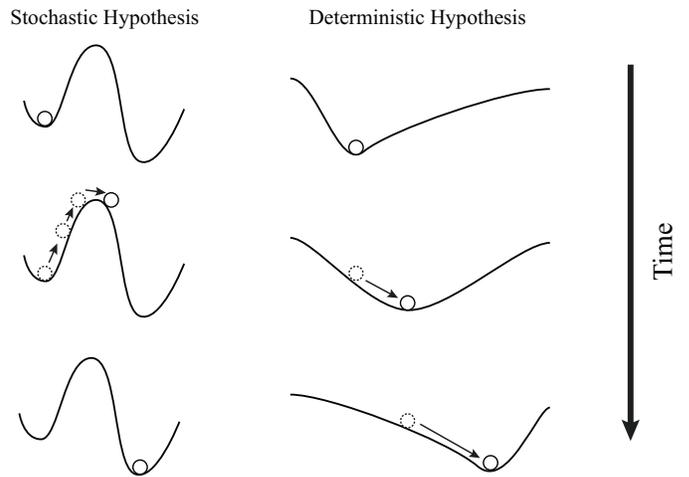}
}
 \caption{Two hypotheses about cell reprogramming mechanism and the basic topology. The essential difference of these two hypotheses is whether the landscape itself changes during reprogramming. In the stochastic hypothesis, the landscape does not change but the state of cell stochastically transits from one steady state to another. In the deterministic hypothesis, the landscape itself change and the state of cell passively moves from one steady state to another.}
 \label{fig:1}       
\end{figure}

 Under this scenario, the stochastic hypothesis represents that the cell fate decision network is a traditional Waddington's landscape, consisting numerous valleys and hills. The ball (cell fate), is able to stochastically transit from one valley to another. Here each valley represents a cell fate and reprogramming process is to trigger cell from one fate to another. On the other hand, in the deterministic hypothesis stochasticity plays a trivial role, there is no transition between two steady state and the cell fate transition is determined by the continuous alternation of the landscape (fig.\ref{fig:1}).

 Therefore, a one key difference between these two models is the reprogramming time in stochastic hypothesis depends not only on the 'energy barrier' between two steady state but also on the magnitude of noise while the reprogramming time in deterministic hypothesis depends only on the dynamics shifting the landscape. This idea is intuitively obvious. However, the quantitative analysis of the timescale required for reprogramming still lacks, especially the contribution of noise to the reprogramming time. If the noise works trivially for the total time for reprogramming, alternating noise magnitude may not lead to any significant change of reprogramming time.

 To calculate the time required for reprogramming of two hypotheses, we want to use one unified model to compare the two hypotheses. According to previous studies on lineage specification, mutual inhibition paradigm is shown to be the most significant concept.\cite{Zon2008}\cite{Huang2011} For example, two opposing transcription factors, GATA1 and PU.1 plays dominant role in erythroid/megakaryocyte and myelomonocytic lineage determination respectively. GATA1 and PU.1 mutual inhibited each other, and also control their own expression, forming an self-activation loop (fig.\ref{Fig:bifurcation}a) \cite{Graf2009}. We will show that this simple model fits both stochastic hypothesis and deterministic hypothesis thus it provides a unified framework for us to compare the two distinct hypotheses.

\section{Methods and Model}
\label{theory}
  As previous shown, the core network which in responsible for the cell fate decision is able to be generally regarded as mutual inhibition \cite{Enver}\cite{M11}: two opposing transcription factors $x$ and $y$ inhibited each other while activated themselves respectively. This topology is able to be described by the following ordinary differential equations:

  \begin{eqnarray}
  \label{eq:one}
    \frac{dx}{dt}=\frac{\alpha_{1} x^{m}}{K^{m}_{xx}+x^{m}} + \frac{\beta_{1}}{K^{m}_{yx}+y^{m}}-k_{1}x\\
    \frac{dy}{dt}=\frac{\alpha_{2} y^{n}}{K^{n}_{yy}+y^{n}} + \frac{\beta_{2}}{K^{n}_{xy}+x^{n}}-k_{2}y
  \end{eqnarray}

\begin{figure*}
  \resizebox{0.75\textwidth}{!}{%
    \includegraphics{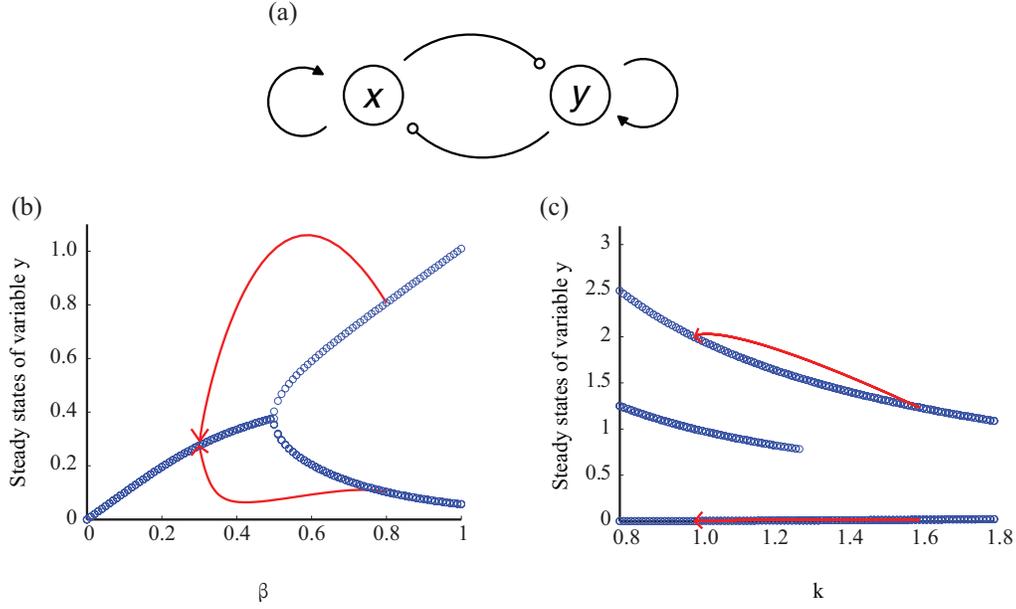}
  }
  \centering
  \caption{\label{Fig:bifurcation}(a): The illustration of mutual inhibition. Two genes inhibit each other while activate themselves respectively. (b): Type I bifurcation ($\alpha=\alpha_{1}=\alpha_{2}=0.01, K=0.5, S = 0.5, n=4, k=k_{1}=k_{2}=1$). Cell differentiation is similar to the process of bifurcation. A control parameter (which in this case is the inhibition coefficient $\beta$) alteration leads to the steady states change and the cell reprogramming process is in reverse. In this scenario, the cell reprogramming is a deterministic process which controlled by a certain parameter. The red line is a typical trajectory of steady state transition led by bifurcation with speed $r = 0.7$ (See eq.\ref{eq:theta}) and the arrow represents the direction. (c): Type II bifurcation ($\alpha=\alpha_{1}=\alpha_{2}=1, \beta=\beta_{1}=\beta_{2}=1.1, K=0.5, S = 0.5, n=4$). A control parameter turn the system with 3 steady states to that with 2 steady states, which may be elucidated as the cell differentiation process. However, in this scenario, cell reprogramming requires not only the parameter alteration but the stochastic transition as well since all the three steady states have no point of intersection. The red line is a typical trajectory of steady state transition led by bifurcation with speed $r=1.2$.}
  \end{figure*}

  These equations are based on Hill functions which are generally used in describing gene expression regulation.\cite{Uri2006} The first item represents the self-activation of each transcription factor; the second item represents the mutual inhibition; and the third item describes the unregulated degradation.
  One of the benefit to use this simple model is that it exists two types of bifurcations (fig.\ref{Fig:bifurcation}). In the type I bifurcation, as the parameter grows the steady state divides into two and the system becomes bistable (supercritical bifurcation). In the type II bifurcation, as the parameter grows, one steady state disappears and the other two change continuously (subcritical bifurcation).

  In the deterministic hypothesis, the stochastic transition contributes trivially, which means the steady state has a deterministic trajectory during bifurcation. In the stochastic hypothesis, both the stem cell and the differentiated cell steady states are intrinsically existed. The cell is able to 'jump' from one steady state to another stochastically. To calculate the reprogramming time, it's necessary to calculate the height of the 'energy barrier' between two steady states. For example, consider a nonlinear dynamic system which has more than one steady states£º

  \begin{equation}
    \frac{d\textbf{x}}{dt}=\textbf{F}(\textbf{x})
  \end{equation}

  which the driving force $\textbf{F}$ exhibits in a nonlinear manner. If the dynamic system is of one dimension, there is always a potential function $U$ satisfied $U_{AB}=\int_{x_{A}}^{x_{B}} F(x)dx$, which means $U$ is the primitive of $F(x)$. In this situation, the transition probability between two different steady states $x_{A}$ and $x_{B}$ is:

  \begin{equation}
    P_{x_{A}\rightarrow x_{B}}=e^{-\frac{U_{AS}}{\varepsilon ^2}}
  \end{equation}

  where $U_{AS}$ is the $\Delta U$ between $x_{A}$ and the highest energy point between $x_{A}$ and $x_{B}$ (the energy barrier). $\varepsilon$ is the magnitude of noise.

  However, when the dimension of the system is greater than 1, the upper theory may not be able to be applied since the high-dimensional, non-equilibrium systems generally are not gradient system, which means the following equation may not be held:

  \begin{equation}
    \frac{d\textbf{x}}{dt}=\textbf{F}(\textbf{x})=-\nabla U
  \end{equation}

  Because only the gradient part of the driving force $\textbf{F}(\textbf{x})$ determines the transition rate between two steady states, it's ideal to decompose the driving force into two parts: a gradient of potential $-\nabla U$ and the remanent as following:

  \begin{equation}
    \frac{d\textbf{x}}{dt}=\textbf{F}(\textbf{x})=-\textbf{D}\nabla U + \textbf{F}_{r}
  \end{equation}

  where $\textbf{D}$ is the diffusion coefficient tensor and here accounting for the magnitude of noise.\cite{M3} There are numerous methods to decompose the driving force\cite{M2}, here we use the decomposition based on the flux of the probability which the probability can be described as a diffusion equation:

  \begin{equation}
    \frac{\partial P}{\partial t} + \nabla \cdot \textbf{J}(\textbf{x},t)=0
  \end{equation}

  The diffusion equation describes a conservation of probability (local change is due to net flux). Flux vector $\textbf{J}$ is defined as $\textbf{J}=\textbf{F}P-\textbf{D}\cdot \frac{\partial P}{\partial\textbf{x}}$.\cite{M4} representing the speed of the flow of the probability in concentration space $\textbf{x}$.

  In the steady state $\frac{\partial P}{\partial t}=0$, then $\nabla \cdot \textbf{J}(\textbf{x},t)=0$. If the system is an equilibrium system, $\nabla \cdot \textbf{J}(\textbf{x},t)=0$ always leads to $\textbf{J}=0$. However, for a nonequilibrium system, this condition does not mean that $\textbf{J}$ have to vanish because the detailed balance condition is satisfied. Therefore, deviating from the definition of $\textbf{J}$, the flux in the steady state is defined as: $\textbf{J}_{ss}=\textbf{F}P_{ss}-\textbf{D}\cdot \frac{\partial P_{ss}}{\partial\textbf{x}}$, so:

  \begin{equation}
    \textbf{F}=\textbf{D}\cdot\frac{\partial P_{ss}}{\partial\textbf{x}}/P_{ss}+\textbf{J}_{ss}/P_{ss}=\textbf{D}\cdot\frac{\partial U}{\partial\textbf{x}}+\textbf{J}_{ss}/P_{ss}
  \end{equation}

  where $U=-lnP_{ss}$. So $U$ reflects the potential and $\textbf{F}_{r}=\textbf{J}_{ss}/P_{ss}$.

  The quasi-potential $U$ is able to be calculated by $U=-lnP_{ss}$ which $P_{ss}$ is calculated through a Fokker-Planck equation:
  \begin{equation}
  \label{eq:Fokker}
    \frac{\partial P}{\partial t}=-\nabla\cdot(\textbf{F}P)+\textbf{D}\nabla^{2}P
  \end{equation}

  This equation was solved numerically by finite difference method with the second boundary condition.(fig.\ref{Fig:quasi})

\section{Results}
\label{Results}
As previous discussed, there are two types of bifurcation generally in the model (see Methods). The first type is illustrated in fig.\ref{Fig:bifurcation}b, which the bifurcation is controlled by the inhibition coefficients $\beta$. This bifurcation could be elucidated from a biological perspective as a deterministic transition from one state (\textit{e.g.} ESC state) to another state (\textit{e.g.} stomatic cell state). The second type of bifurcation is illustrated in fig.\ref{Fig:bifurcation}c, which the degradation rate $k$ controls the systems with 3 steady states to 2 steady states. This type may be elucidated as the ESC state goes to the differentiated state like stomatic cell through bifurcation or stochastic transition from one steady state to another and reprogramming required stochastic state transition but not only bifurcation.

Our aim is to discriminate the two hypotheses by their characteristic time for reprogramming. To simulate the experiments approach of reprogramming, we used a simple exponential equation to represent the effect of transformation of genes into somatic cell to reprogramme\cite{I4} or add chemical small molecules to reprogramme\cite{R1}, which trigger the control parameter's change.
 \begin{equation}
 \label{eq:theta}
    \theta_(t)=\theta_{b}+(\theta_{0}-\theta_{b})e^{-rt}
  \end{equation}
  Here $\theta$ is the control parameter which in the type I bifurcation is $\beta$, and in the type II bifurcation, $k$. $\theta_{b}$ is the basal level of $\theta$, which means the final $\theta$ value; $\theta_{0}$ is the original level of $\theta$, representing the $\theta$ value of differentiated cell state; and $r$ represents how fast $\theta$ declines. When the system with $\theta_{0}$ reaches its steady state, $\theta$ begins to change as in Equation \ref{eq:theta}, then the time for reaching the new steady state is calculated.

 \begin{figure*}
   \resizebox{0.75\textwidth}{!}{%
   \includegraphics{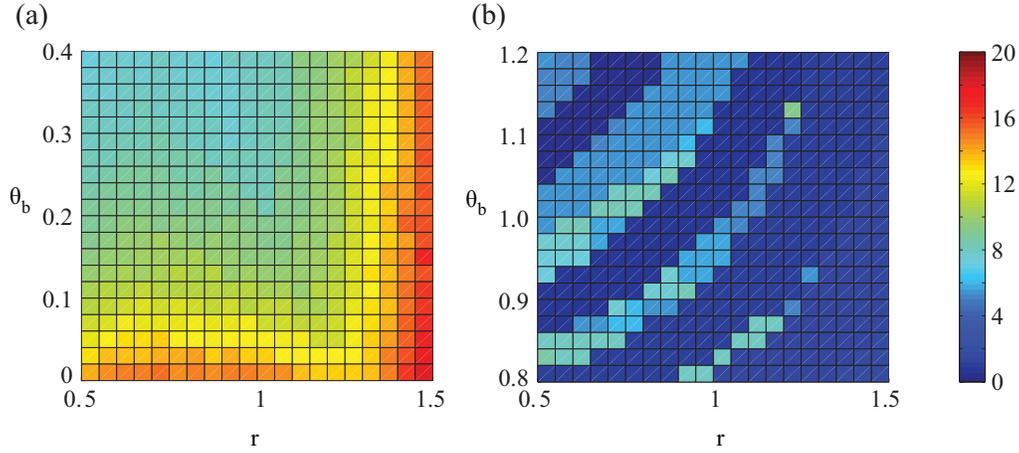}
   }
   \centering
   \caption{\label{Fig:timescale}Timescales required for steady states transition during bifurcation. a: Type I bifurcation. The time is the total time for cell reprogramming. b: Type II bifurcation. The time is only the first part of the total time for reprogramming. It shows that in a wide range of declining rate $r$ and $\theta_{b}$, the time scale is $10^{1}$. The parameter used in this paper is normalized by the dimension unit of protein degradation rate, which in mammalian cell is usually about $10^{-1}h^{-1}$, so the time calculated here would be 10 times greater than the actual time. In the rest of the paper, we used this non-dimensional-normalized time unit.}
 \end{figure*}

\subsection{Deterministic Hypothesis}
\label{Deter}
  The time for reprogramming in the type I bifurcation is calculated through the method above. In this scenario, the time to reaching the new steady state is exactly the time for reprogramming since there is no stochastic transition. A typical trajectory of reprogramming is shown in fig.\ref{Fig:bifurcation}b (red line, the arrow indicates the direction). The results (fig.\ref{Fig:timescale}a) shows that in a wide range of declining rate $r$ and the final $\theta$ value, the timescale for reprogramming is $10^{1}$.

\subsection{Stochastic Hypothesis}
\label{Sto}
\begin{figure*}
    \resizebox{0.75\textwidth}{!}{%
    \includegraphics{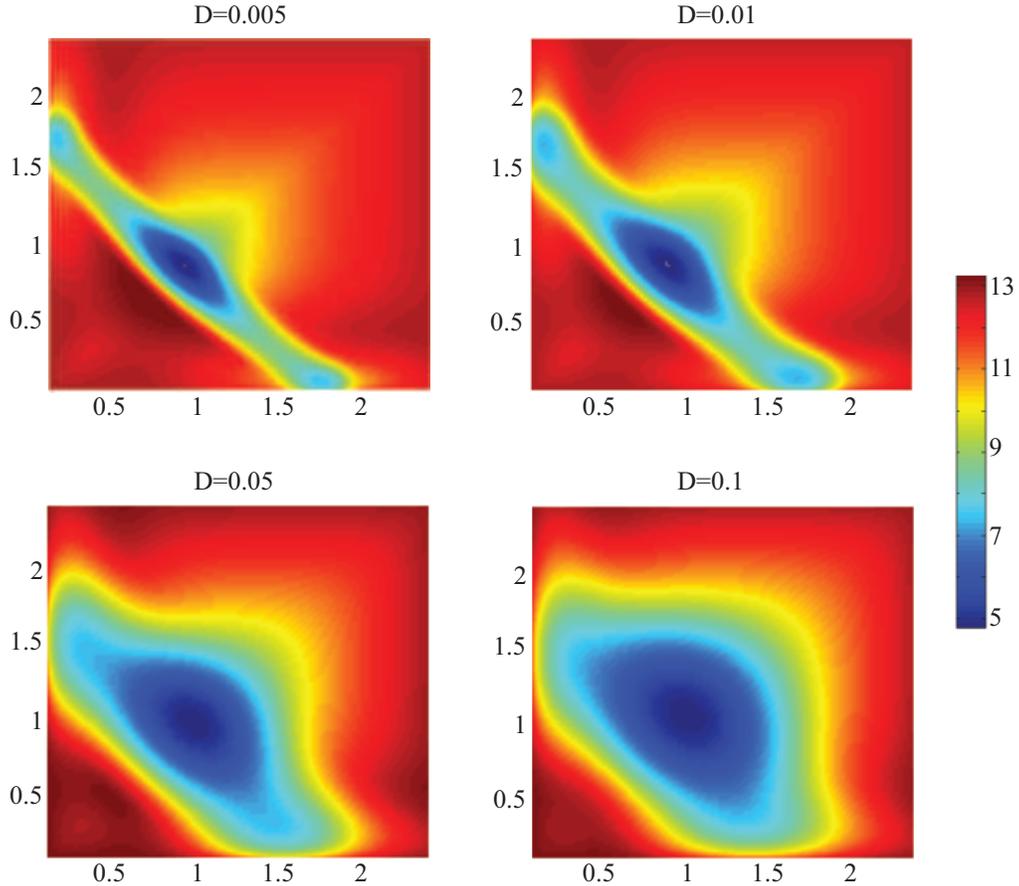}
    }
    \centering
    \caption{The Quasi-potential landscapes with different noise magnitudes. $\alpha=\alpha_{1}=\alpha_{2}=1, \beta=\beta_{1}=\beta_{2}=1.1, K=0.5, S=0.5, n=4, k=k_{1}=k_{2}=1.2$. $U=-lnP_{ss}$ is the quasi-potential as described ahead (See Methods). The quasi-potential was obtained by the finite difference method\cite{M2}, $P^{k+1}=P^{k}+\tau\cdot\delta$. After the iteration, the sum of $\delta$ is less than $10^{-6}$. It indicates that when noise is larger than 0.05, there is only one attractor, which means cells will not stay in the differentiated states anymore in a very noisy environment. However, if the noise magnitude is $10^{-3}\sim10^{-2}$ as analyzed in the paper, there will remain three steady states and the stochastic transition is necessary.}
    \label{Fig:quasi}
  \end{figure*}

  The time for reprogramming in the type II bifurcation consists of two parts: the deterministic transition and stochastic transition. The first part is calculated as the above-mentioned approach. The results is shown in fig.\ref{Fig:timescale}b, which indicated that the characteristic timescale is about $10^{0}$ to $10^{1}$. A typical trajectory is shown in fig.\ref{Fig:bifurcation}c (red line), it demonstrates that without the stochasticity, the differentiated states can never return to the stem cell state. For the second part, the Fokker-Planck equation (see Methods, eq.\ref{eq:Fokker}) is solved (fig.\ref{Fig:quasi}).

\begin{figure*}
    \resizebox{0.75\textwidth}{!}{%
    \includegraphics{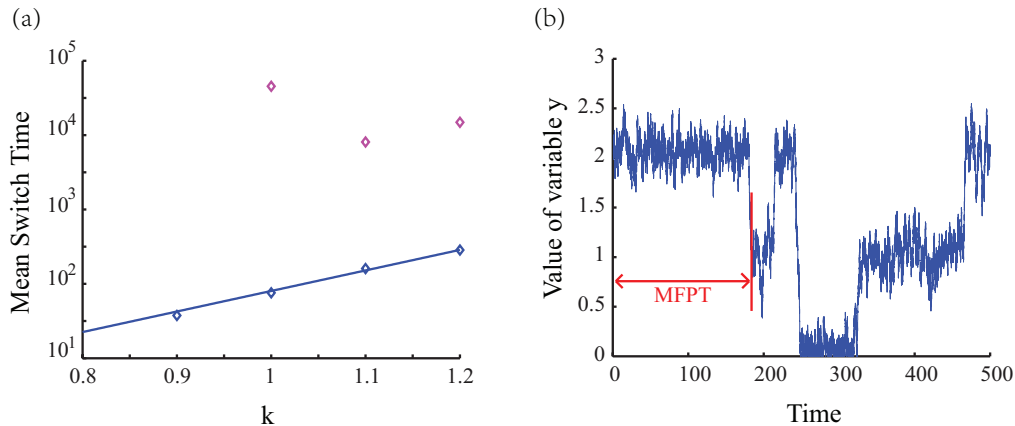}
    }
    \centering
    \caption{(a): The Mean Switch Time under different noise magnitudes. Blue dots are the MST with $D=0.01$, and blue line is the fitting results from the Langevin equation. Purple dots are MST with $D=0.005$. It shows that a slight diminishment of noise leads to a tremendous change of MST. Nevertheless, time of stochastic transition cannot be neglected in the \textit{in vivo} biological noise magnitude. (b): Using stochastic simulation to estimate MST and the prefactor $T$, here the MST is estimated by the Mean First-Passage Time (MFPT). The simulation begins in the steady state, and the MFPT is the time when the variable first pass the other steady state determined by ODE. Here $T$ is fitted as 0.1386.}
    \label{Fig:MST}
  \end{figure*}

  The large deviation theory predicts when a rare event occurs, it will follow the optimal transition path with high probability.\cite{Varadhan} So the energy barrier is the quasi-potential difference between the steady state and the saddle point between two steady state (fig.\ref{Fig:quasi}). According to \cite{Kampen}\cite{Keizer}, the noise magnitude $D$ of Fokker-Plank equation is approximately equal to $\frac{1}{2}\Omega^{-\frac{1}{2}}$, which $\Omega$ is the system's size, that here is chosen as the average number of molecules. If so, since most proteins in a mammalian cell have average numbers about $10^{4}\sim 10^{9}$ magnitude,\cite{Schwan}\cite{Nagaraj} the noise magnitude $D$ here would be about $10^{-3}\sim10^{-2}$. We calculated the energy barrier with control parameter $k$ value of 1.0~1.2 and different noise magnitudes $D$ ranging from 0.005 to 0.1. The results are shown in fig.\ref{Fig:MST}.

  Then the time for stochastic transition is able to estimate according to \cite{Wentzell}. The Mean Switch Time (MST) $\tau$ of transition from differentiated state to stem cell state is:

  \begin{equation}
  \label{eq:MST}
    \tau\approx T e^{K\Delta S}
  \end{equation}

  Here $T$ is a prefactor, $K$ is the system size which is usually chosen as the typical number of proteins, and $\Delta S$ is the energy barrier. However, there are no available methods to obtain $T$ analytically in high dimensions for the geometry problem and non-gradient nature.\cite{Stein} Fortunately, $T$ varies slowly in many cases \cite{Lv}, therefore it can be fitted from the numerical results. We apply the stochastic simulation by a Langevin approach and the MST is estimated by Mean First-Passage Time (MFPT).

  The results are shown in Fig.\ref{Fig:MST}, it indicates that noise in a reasonable magnitude plays an strikingly significant role. The average time for stochastic transition is in the timescale about $10^2$, which is much longer than the time of steady steady state transition led by bifurcation (Fig.\ref{Fig:timescale}). However, if the noise can be tuned at least 10 times lower than natural level, the time of stochastic transition is negligible.

  Surprisingly, the relationship between MST (or energy barrier $\Delta S$) and $k$ is depended on the noise magnitude $D$. When the noise is in a medium strength ($D=0.01$), MST is increasing exponentially with $k$. However, when the noise decreases ($D=0.005$), MST has a minimum value when $k=1.1$ (fig.\ref{Fig:MST}). The results indicate that the noise influence not only the Mean Switch Time, but also the optimum value of the control parameter $k$.

\section{Discussion and Conclusion}
\label{discussion}
\subsection{Biological Significance}
\label{significance}
  In this paper, we have presented a general framework of cell reprogramming and proposed a new methodology to test the nature of reprogramming process. Our results indicate that if reprogramming is a stochastic process, the magnitude of noise influences significantly the total time required for reprogramming. However, so far there is few experiments focus on the role of noise in the cell reprogramming.

  Biological noise has been studied intensely in recent years, we know for a constitutive gene, the noise strength is:\cite{Ouden}

  \begin{equation}
  \frac{\sigma^{2}_{p}}{<p>}\cong 1+b
  \end{equation}

  Here $b$ is the average number of proteins translated by per mRNA transcript. Therefore it's possible to tune the noise magnitude without altering the average gene expression level. This idea is able to test experimentally. For example, in \cite{I41}, the cells are reprogrammed by exogenous gene expressions. If the stochastic hypothesis holds, the time for reprogramming would be significantly different in different magnitudes of noise of those exogenous gene expressions. Our data indicates if the noise strength was tuned to 10 times greater than the current magnitude, the cell for reprogramming would reduce more than 10 times. If the deterministic hypothesis holds, the time would not change. Tuning the noise will not change the time for reprogramming but the percentage of iPS cell because the fluctuation of protein level.

  Another role that noise may facilitate research of cell reprogramming is fluctuation correlation. In the type I bifurcation, the reprogramming process experiences a supercritical bifurcation, which the noise may show a fluctuation correlations \cite{Scheffer}, a critical phenomenon in statistical physics. However, in the type II bifurcation, the transition of steady state does not pass trough the bifurcation point so the fluctuation correlation would not occur. So far there is no experiments applied focusing on this prediction and it would be interesting to investigate both theoretically and experimentally.

\subsection{Limitations}
\label{limit}
  Our results, based on dynamical systems and perturbation theory, provide a theoretical framework which may facilitate the rational design of cell reprogramming protocols in the future. We focus on the temporal character, the timescale, but not the properties of steady states. However, our results are based on a minimum mathematical model with artificial parameters. The real biological network for cell-fate decision may be much more complex, and in different systems, the network may be also different. For example, in erythroid/megakaryocyte and myelomonocytic lineage determination mentioned in this paper, mutual inhibition with self-activation is the core network, while the embryonic stem cell-fate decision depend on a "Seesaw" Model \cite{Shu}. Besides, the parameters used in the model is regarded as symmetrical (\textit{e.g.} $\beta_{1}=\beta_{2}, k_{1}=k_{2}$), when parameters are altered unsymmetrically, the bifurcation diagram will be slightly different. Our model is a kind of "toy model" captured only the essential features, the application to a real biological network deserves future work.

  In summary, our model provides a novel perspective to test the hypotheses of cell reprogramming. We studied a general model which is able to fit both deterministic and stochastic hypothesis and quantitatively calculated the time to re-reaching the steady state during bifurcation and mean switch time of stochastic transition. We hope our results may facilitate the future research on cell reprogramming.

\end{document}